# Unlocking the out-of-plane dimension for photonic bound states in the continuum to achieve maximum optical chirality

Lucca Kühner[1,†], Fedja J. Wendisch[1,†], Alexander A. Antonov[2], Johannes Bürger[1], Ludwig Hüttenhofer[1], Leonardo de S. Menezes[1,3], Stefan A. Maier[4,1,5], Maxim V. Gorkunov[2], Yuri Kivshar[6], and Andreas Tittl[1,*]

[1]Chair in Hybrid Nanosystems, Nanoinstitute Munich, and Center for NanoScience, Faculty of Physics, Ludwig-Maximilians-University Munich, Königinstrasse 10, 80539 München, Germany

[2]Shubnikov Institute of Crystallography, FSRC "Crystallography and Photonics", Russian Academy of Sciences, Moscow 119333, Russia

[3]Departamento de Física, Universidade Federal de Pernambuco, 50670-901 Recife, Pernambuco, Brazil

[4]School of Physics and Astronomy, Monash University, Wellington Rd, Clayton VIC 3800, Australia

[5]The Blackett Laboratory, Department of Physics, Imperial College London, London, SW7 2AZ, United Kingdom

[6]Nonlinear Physics Centre, Research School of Physics, Australian National University, Canberra ACT 2601, Australia

[†]These authors contributed equally

*e-mail: andreas.tittl@physik.uni-muenchen.de

**Abstract**

The realization of lossless metasurfaces with true chirality crucially requires the fabrication of three-dimensional structures, constraining their feasibility for experiments and hampering practical implementations. Even though the three-dimensional assembly of metallic nanostructures has been demonstrated previously, the resulting plasmonic resonances suffer from high intrinsic and radiative losses. The concept of photonic bound states in the continuum (BICs) is instrumental for tailoring radiative losses in diverse geometries, especially when implemented using lossless dielectrics, but applications have so far been limited to planar and intrinsically achiral structures. Here, we introduce a novel nanofabrication approach to unlock the height of generally flat all-dielectric metasurfaces as an accessible parameter for efficient resonance and functionality control. In particular, we realize out-of-plane symmetry breaking in quasi-BIC metasurfaces and leverage this design degree of freedom to demonstrate, for the first time, an optical all-dielectric quasi-BIC metasurface with maximum intrinsic chirality that responds selectively to light of a particular circular polarization depending on the structural handedness. Our experimental results not only open a new paradigm for all-dielectric BICs and chiral nanophotonics but also promise advances in the realization of efficient generation of optical angular momentum, holographic metasurfaces, and parity-time symmetry-broken optical systems.

Controlling the interaction of different polarization states of light with matter has been a fundamental aim of optics, covering the gamut from fundamental science[1] to practical technological applications[2]. Metasurfaces composed of resonant subwavelength building blocks with tailored optical properties have significantly advanced the capabilities for controlling light on the nanoscale[3], launching breakthrough applications in diverse fields including localized high-harmonic generation[4,5], ultrathin optical elements[6,7], and biomolecular sensing[8–10]. In recent years, two major developments have underpinned the rapid progress in the metasurface technology and applications: (i) a transition from traditional plasmonic resonator geometries to all-dielectric materials to overcome Ohmic losses, and (ii) the utilization of the emerging physical concept of photonic quasi bound states in the continuum (qBICs), which provides versatile control over the radiative losses in nanophotonic systems[11–14]. Combining these advances, all-dielectric qBIC-driven metasurfaces have delivered ultrasharp resonances with high values of the quality factors ($Q$ factors)[4,15], broad spectral tunability[9,16], and strongly enhanced near-fields for boosting both surface-driven and material-intrinsic processes[5,17,18]. Among different BIC-driven concepts[19], metasurfaces with broken in-plane inversion symmetry are especially appealing for tailoring light-matter coupling, because they enable a straightforward control over the radiative lifetimes via geometric perturbations within the meta unit[12].

So far, most qBIC-driven metasurface realizations have relied on modifying the in-plane geometry of the resonant elements to control the asymmetry, owing to the challenges of fabricating resonators with different heights at subwavelength distances. This limitation also constitutes a significant roadblock for applications such as holography[20–22], generation of the beams with an optical angular momentum (OAM)[23], chirality sensing[24], and chiral nanophotonics[25–27], which naturally require non-planar structures to enable the efficient interaction with more complex polarization states of light. In particular, chiral qBICs have recently been proposed theoretically[28–30], aiming to provide efficient coupling with circularly polarized light and circular dichroism responses with greatly reduced linewidths. However, proof-of-concept experimental implementations remain limited to the microwave range[31], while optical realizations faced severe restrictions associated with complex three-dimensional unit-cell designs.[26,31,32]

Here, we experimentally demonstrate out-of-plane symmetry-broken qBIC metasurfaces in the red part of the visible spectrum by leveraging a novel multi-step fabrication process for arbitrary height control of the resonators. Crucially, our approach implements different resonator heights based on an additional deposition step of the dielectric material, allowing height control with extreme precision down to the Angstrom range, only limited by the parameters of the respective evaporation or sputtering processes. We first utilize this approach to realize height-driven qBIC resonances with tailored linewidths interacting with linearly polarized light, and then we generalize this concept to demonstrate chiral qBIC metasurfaces that selectively couple to circularly polarized light depending on the structural handedness. Our results and fabrication method relax the constraints of purely planar metasurface geometries, and thus offer a nontrivial generalization to the entire metasurface concept unlocking an additional degree of freedom and extending independent parameters for freely tuning the optical response of metasurfaces, significantly increasing their design flexibility and delivering previously unavailable functionalities requiring multi-height geometries.

**Results and discussion**

**Out-of-plane qBIC engineering**

The optical properties of qBIC metasurfaces can be modeled using coupled-mode theory (CMT) describing light scattering as an interference of a direct background channel and a resonant channel underpinned by a qBIC eigenstate excitation and re-radiation[33]. The eigenstate interaction with the far field is described by the coupling parameters $m_\mathbf{e}$ and for electromagnetic waves normally incident along the *z*-axis and polarized along unit vectors **e** they are proportional to[12]:

$$m_\mathbf{e} \propto \int_V \mathbf{J}(\mathbf{r}) \cdot \mathbf{e} \; e^{ikz} \, dV, \tag{1}$$

where **J(r)** is the displacement current within the meta unit volume *V* and *k* is the light wavenumber. The coupling coefficients $m'_\mathbf{e}$ to waves incident onto the metasurface backside are expressed by similar integrals with reversed propagation direction ($e^{ikz} \rightarrow e^{-ikz}$). The corresponding power transmission coefficients of an incident wave (polarized along unit vector **i**) into an outgoing wave (polarized along unit vector **f**) are expressed as:

$$T_\mathbf{fi}(\omega) = |t_\mathbf{fi}(\omega)|^2, \text{ with } t_\mathbf{fi}(\omega) = \tau \delta_\mathbf{fi} - \frac{m_\mathbf{i} m'_\mathbf{f}}{i(\omega - \omega_0) - \gamma_0}, \tag{2}$$

where $\tau$ is a coefficient of background transmission preserving the polarization ($\delta_\mathbf{fi}$ is the Kronecker delta-symbol), and the complex eigenfrequency ($\omega_0 + i\gamma_0$) contains an imaginary part with radiative and dissipative contributions: $\gamma_0 = \gamma_\mathrm{r} + \gamma_\mathrm{d}$.

QBICs with versatile polarization properties can be realized starting from a simple symmetry-protected antiparallel electric dipole BIC of a double-rod meta unit shown in Fig. 1a. The corresponding coupling parameters given by Equation (1) are reduced to:

$$m_\mathbf{e} \propto \mathbf{p}_1 \cdot \mathbf{e} \; e^{ikz_1} + \mathbf{p}_2 \cdot \mathbf{e} \; e^{ikz_2}, \tag{3}$$

where $\mathbf{p}_{1,2}$ are the electric dipole moments of the rods 1 and 2, and $z_{1,2}$ are their effective z-coordinates. Perfect BIC isolation with $m_\mathbf{e} = 0$ for all polarization unit vectors **e** is achieved when $\mathbf{p}_1 = -\mathbf{p}_2$ and $z_1 = z_2$, i.e., when identical rods are placed within the same plane.

An exemplary in-plane symmetry breaking, transforming this BIC into a qBIC, occurs when the length of one rod is varied and the rod length difference ΔL becomes the asymmetry parameter (Fig. 1a). For ΔL > 0, the electric dipole moments remain antiparallel, but have different magnitudes ($|\mathbf{p}_1| \neq |\mathbf{p}_2|$), which enables the coupling to waves linearly polarized along the long rod axis. Crucially, such coupling can also be tailored via the rod height difference $\Delta h$. In this case, both the magnitudes of the dipole moments and their effective locations become slightly different. Thus, the coupling to the far field is enabled by $m_y \propto p_1 (e^{ik\Delta z} - 1) + \Delta p$, where both $\Delta p = p_1 - p_2$, and $\Delta z = z_1 - z_2$ are determined by $\Delta h$ as it produces a difference in the rod volume and also shifts their centers of mass, see Fig. 1b. In all above cases, the rods remain parallel and the qBIC is linearly polarized contributing solely to $T_{yy}$.

Combining height-driven out-of-plane symmetry breaking perturbation with a conventional in-plane one enables realizing chiral qBIC metasurfaces[29,31]. Specifically, we implement a metasurface that breaks all point symmetries by utilizing a meta unit composed of two rods of equal lengths diverged in-plane by a small angle $\theta$ and having different footprints and heights while maintaining identical cross sections (Fig. 1e). In this case, the electric dipoles $\mathbf{p}_1$ and $\mathbf{p}_2$, even though equal in magnitude ($|\mathbf{p}_1| = |\mathbf{p}_2|$), are not exactly antiparallel, and their effective z-coordinates also differ, producing a three-dimensional chiral arrangement. It is insightful to evaluate Equation (3) in such a case and obtain for the coupling parameters to the left circularly polarized (LCP) and right circularly polarized (RCP) waves:

$$m_{L,R} \propto \sin(\theta \pm k\Delta z/2). \qquad (4)$$

Precise tailoring of the meta unit geometry to balance the perturbations via $2\theta = k\Delta z$ enables efficient and controllable qBIC coupling to the LCP waves with $m_L \propto \sin(2\theta)$, whereas sustaining full qBIC isolation from the RCP waves by $m_R = 0$. Notably, according to Equation (2), an ideally matched chiral metasurface does not convert circular polarizations from LCP to RCP ($t_{RL} = 0$) or from RCP to LCP ($t_{LR} = 0$), although there are no symmetry restrictions which, for example, forbid such conversions in the presence of rotational symmetry axes[34].

Remarkably, this simple design allows approaching the ultimate limit of maximum chirality[35], when an object remains transparent to the waves of one circular polarization, for instance RCP, and strongly interacts with those of the opposite polarization (LCP). To realize this, one has to first ensure that the qBIC resonance occurs in the spectral range of full background transparency ($|\tau| \approx 1$) and negligible dissipation in the metasurface material ($\gamma_0 = \gamma_r$). Then the coupling coefficients to LCP light incident onto the metasurface front and back satisfy $m'_L = -m_L^*$ and the decay rate $\gamma_r$ follows a characteristic quadratic dependence on the coupling parameter $\gamma_r = |m_L|^2$, leading to zero transmittance $t_{LL}$ at resonance[31] ($\omega = \omega_0$) given by Equation (2).

Note that maximum chiral metasurfaces qualitatively outperform those exhibiting bands of similarly strong circular dichroism (CD), such as plasmonic chiral hole arrays[36]. For the latter, the CD reaches its extreme $\pm 1$ values when one circular polarization is fully blocked regardless of the transmission of the opposite one. Strong CD of maximum chiral qBIC metasurfaces, on the contrary, is achieved when a selective blocking of waves of one circular polarization is accompanied by close to unitary transmission of their counterparts. As a specific parameter quantifying such exceptional transmission selectivity, we introduce the transmittance difference:

$$\Delta T = T_{RR} - T_{LL}, \tag{5}$$

to specifically characterize the proximity to maximum chirality: while the conventional $CD = (T_{RR} - T_{LL})/(T_{RR} + T_{LL})$ tends to $\pm 1$ as soon as waves of a certain circular polarization are fully blocked, the difference $\Delta T$ approaches $\pm 1$ only if, additionally, the waves of the opposite polarization are fully transmitted.

Left-handed and right-handed enantiomers of the chiral qBIC metasurface can be realized by swapping the rods (Fig. 1e). The opposite enantiomer similarly remains transparent to LCP waves and resonantly blocks RCP waves. Importantly for applications, in contrast to metasurfaces with rotational symmetry axes[37,38], the blocked circularly polarized light is not absorbed but is rather reflected backwards (see Fig. S1). Therefore, under perfect circumstances, each metasurface enantiomer acts as a maximum chiral lossless filter of the corresponding handedness.

**Numerical simulations and optimization**

To verify the height-driven qBIC engineering, we perform numerical simulations of the transmission and reflection of normally incident light as described in the Methods section. For linearly polarized qBICs, we model square metasurface lattice of a period of 450 nm with the meta unit consisting of parallel rods with equal footprint of 330 × 100 nm² and a base height of 120 nm with a variable height difference $\Delta h$ (Fig. 1a). Simulated transmittance spectra for linearly polarized light along the rod axis are shown in Fig. 1b, revealing strong and sharp qBIC resonances in the red part of the visible spectrum as well as the direct control over the resonance linewidth and modulation afforded by the asymmetry parameter $\Delta h$. As expected for BIC-driven systems, the resonance is absent for the symmetric case ($\Delta h = 0$ nm, black dashed line in Fig. 1b) and starts to couple to the radiation continuum for increasing asymmetry, showing the sharpest resonances with $Q$ factors above $10^4$ for $\Delta h = 5$ nm. To quantify the influence of the asymmetry parameter on the resonance sharpness, we extract the resonance $Q$ factors from the transmittance spectra by fitting them with a coupled mode theory (CMT) model and plotting them as a function of $\Delta h$ (Fig. 1c). We find that the $Q$ factors follow an inverse quadratic relationship with the asymmetry ($Q \propto 1/\Delta h^2$), which is a hallmark feature of qBIC metasurfaces, confirming that our height-driven symmetry breaking approach fits within the established BIC framework. As a further confirmation, we simulate the electric near-field distribution at the resonant wavelength (Fig.1d). The characteristic mode structure of antiparallel dipoles is nicely reproduced in the numerical simulations, and we observe high near-field enhancements $|E/E_0|$ exceeding 50, which is competitive with previous symmetry-breaking approaches[9]. Notably, the Angstrom-level control over the asymmetry provided by state-of-the-art material deposition technologies such as atomic layer deposition (ALD) can enable ultrasmall values of $\Delta h$ and therefore much higher field enhancements (Fig. S2). Combined with the strong confinement of fields to the resonator surface (Fig. 1d) this makes such height-driven geometries ideal candidates for enhancing interface-driven processes such as biospectroscopy or catalysis.

Next, we simulate chiral configurations by utilizing a meta unit consisting of a pair of rectangular rods diverged by a small angle $\theta$ from the y-axis (Fig. 1e). The square lattice period is set to 550 nm and the rod centers are placed equidistantly with a 275 nm spacing between them. To preserve the equality of the electric dipole moment magnitudes ($|\mathbf{p}_1| = |\mathbf{p}_2|$) we sustain identical 160 x 100 nm² cross section of both rods and keep their lengths equal to $L = 310$ but introduce the asymmetry by turning one rod over to its side, as shown in Fig. 1e. For the resulting $\Delta h = 60$ nm, one can roughly estimate the relative displacement of the dipole moments along the z-axis by $\Delta z \approx 30$ nm. Then, for the resonant wavelength around 900 nm, the proportionality $2\theta = k\Delta z$ is fulfilled for a diverging angle $\theta \approx 6°$. For a more precise determination, we perform a series of numerical simulations for different $\theta$ (see Fig. S3) and obtain that $\theta = 8.5°$ corresponds to the optimal transmission of circularly polarized waves shown in Fig. 1f. As illustrated by Figure 1g, smaller angles produce weaker chirality, while larger ones also suppress the maximum values of $\Delta T$.

In the optimal chiral configuration, as envisioned by the CMT phenomenology, the qBIC is fully decoupled from the normally incident RCP waves but gives rise to a pronounced transmission resonance for the LCP waves at a wavelength of 892 nm. The corresponding near-field pattern under LCP illumination (Fig. 1f) demonstrates a maximum enhancement of the local fields by a factor of more than 20.

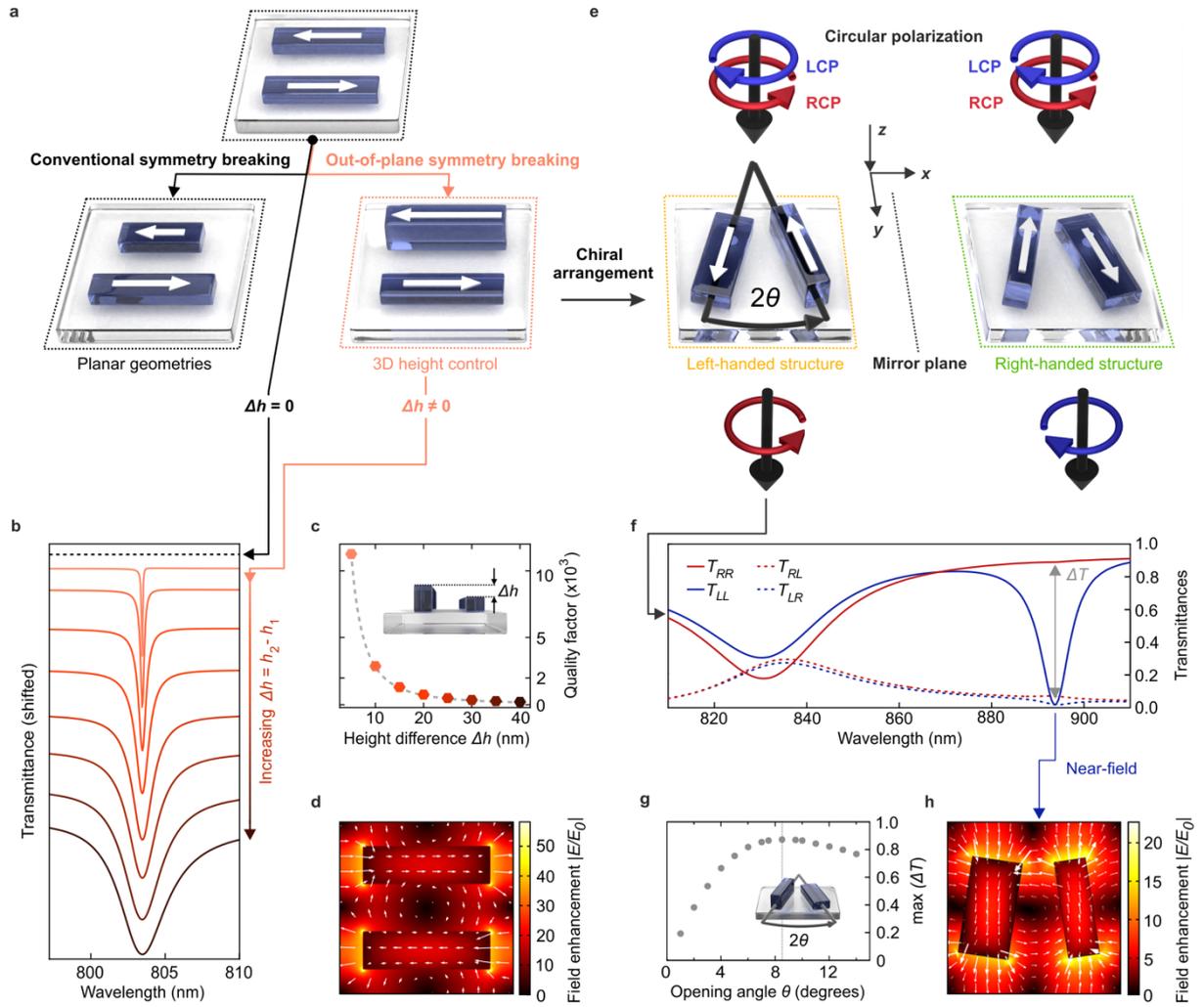

**Fig. 1 | Unlocking the height of dielectric resonators for photonic qBIC engineering.** Established qBIC geometries use in-plane symmetry breaking to couple otherwise dark BIC states to the radiation continuum **(a)**. In our work, we introduce out-of-plane symmetry breaking for qBICs of optical metasurfaces. **b,** Simulated spectral responses for various height differences starting from $\Delta h$ = 0 nm (black dashed line, top) up to $\Delta h$ = 40 nm (dark brown curve, bottom) with the corresponding electric near-field in **(d)**. **c**, Corresponding Q factors extracted from the spectra in **(b)** following the inverse quadratic dependence (fitted as grey dashed line) typical for symmetry-broken qBIC metasurfaces. **e,** Tailoring the height-driven qBIC by an opening angle $\theta$ for the maximum chirality. **f,** Spectral response of a left-handed structure to different incident circular polarizations showing a pronounced qBIC resonance only for LCP light. **g,** Dependence of the maximum $\Delta T$ on the opening angle $\theta$ showing a maximal chiral response at $\theta = 8.5°$. **h,** Corresponding near-field of the chiral qBIC resonance.

**Experimental realization of height-driven BICs**

For the realization of all-dielectric metasurfaces with meta units incorporating resonators of different heights, we demonstrate a new multi-step nanofabrication approach. The core mechanism of this method leverages the combination of an N-step electron beam lithography process and an N-step deposition process to obtain a metasurface with N different height levels. The fabrication steps are illustrated in Fig. 2a for N = 2, as required for our two level height-driven target geometries introduced in Fig. 1. Detailed process parameters are given in the Methods. Importantly, our method is fully scalable for large N, since additional lithography/deposition steps can be added at any point during sample fabrication to obtain additional height levels.

In essence, the metasurface fabrication starts with the plasma-enhanced chemical vapor deposition (PECVD) of an amorphous silicon (a-Si) layer onto a silicon dioxide ($SiO_2$) substrate, where the thickness $h_1$ of the layer defines the height of the resonator element with lowest thickness in the meta unit (Fig. 2a, left). Electron-beam lithography, metal deposition, and wet-chemical lift-off are then performed to obtain a thin chromium (Cr) hard mask defining the footprint of the first resonator (Fig. 2a, step (1)). Subsequently, a second layer of a-Si is deposited onto the sample with a precisely controlled thickness $\Delta h$, which produces a total thickness of $h_2 = h_1 + \Delta h$ for the second resonator element (Fig. 2a, step (2)). A second hard-masking step with accurate spatial alignment is then performed to define the footprint of the second resonator (Fig. 2a, step (3)). At this point, additional pairs of a-Si and hard-masking could be performed to increase the number of height levels of the metasurface. Finally, reactive ion etching is used to transfer the resonator patterns into the a-Si layers (Fig. 2a, step (4)), resulting in pure silicon structures with different height levels after wet-chemical removal of the Cr hard masks (Fig. 2a, step (5)).

We confirm the successful fabrication of the multi-height metasurface structures by atomic force microscopy (AFM) and scanning electron microscopy (SEM). The as-designed metasurface pattern and height differences (Fig. 2b) are already evident from the different scattering intensities in SEM micrograph in Fig. 2c. The SEM image further confirms the good spatial alignment between the lithography steps associated with the two resonator heights (dashed boxes in Fig. 2c, and Fig. S4) although our numerical simulations only show a small impact of the relative rod alignment (Fig. S5) for our alignment accuracy of 5 nm. The AFM measurements reveal accurately defined height differences between adjacent resonators for three different height-driven metasurfaces with asymmetries of 10 nm, 20 nm, and 40 nm (Fig. 2d).

The asymmetry-dependent optical response of the metasurfaces is characterized using confocal white light transmission microscopy (Fig. 2e). In excellent agreement with our simulations, we observe pronounced qBIC resonances when the incident light is polarized along the long rod axis, whereas the resonances disappear for the orthogonal polarization in accordance with the BIC mechanism. Comparing the height-driven metasurface samples for different values of $\Delta h$, we find a clear increase of the resonance $Q$ factor for decreasing asymmetry, highlighting the resonance tailoring capabilities of the method.

From our experimental results, we conclude that our fabrication approach provides a toolkit for realizing multi-element metasurfaces with tailored heights limited only by the precision of the utilized deposition tool, enabling previously unavailable Angstrom-level control over the asymmetry in height-driven BIC geometries.

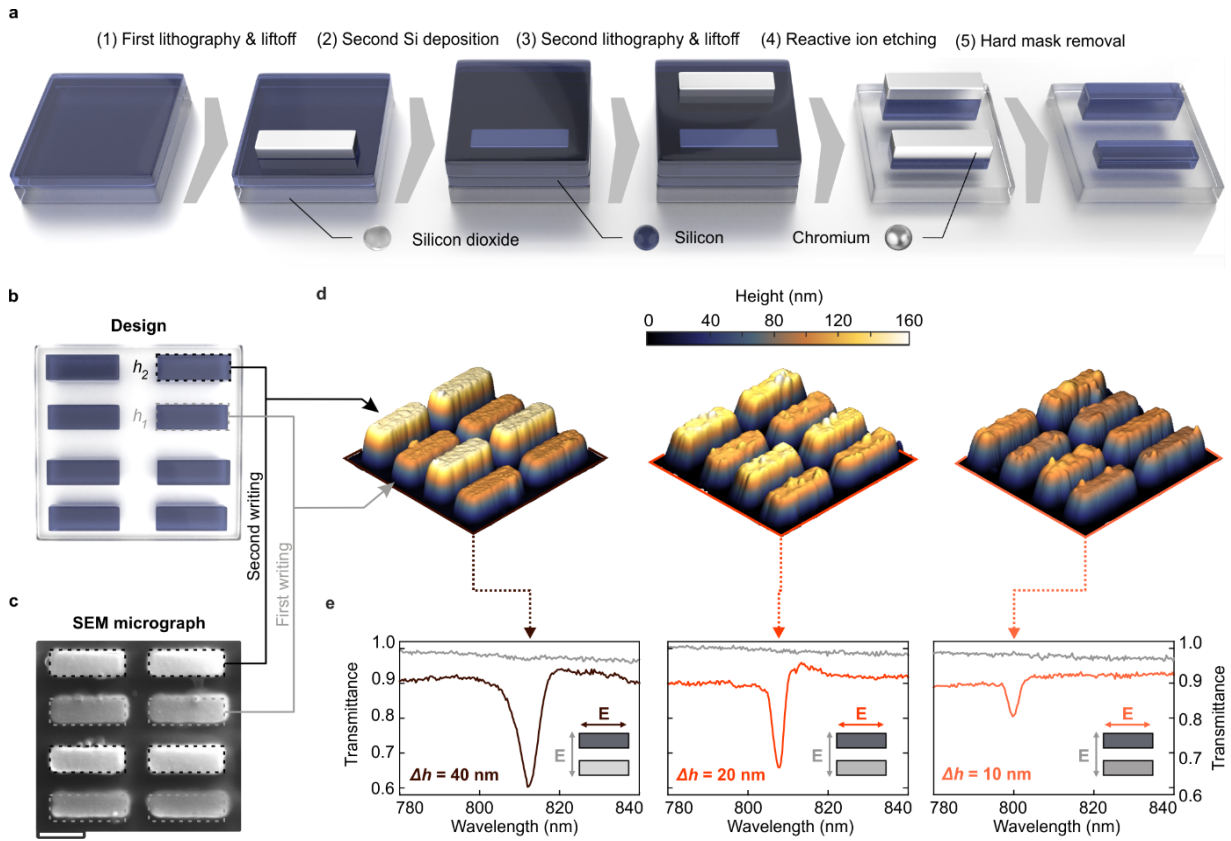

**Fig. 2 | Fabrication principle and precise height control demonstration. a,** Workflow for the fabrication of height-driven metasurfaces. **b,** Schematic illustration of the two-step lithography process along with an SEM micrograph in (**c**) after the fabrication process (scale bar: 200 nm). The height difference Δh is already apparent from the different scattering intensities but also clearly obvious from the 3D AFM micrographs in (**d**). **e,** Optical white light transmittance spectra confirm the precise controllability over the asymmetry Δh via the increase of the respective $Q$ factor as the coupling of the BIC to the radiation continuum decreases. The qBIC resonance is absent for linear polarization of the incident light along the short rod axis (see grey curves in panel (**e**)).

**Height-driven chiral qBICs**

Using our multi-step fabrication approach, we realized left-handed, right-handed, and achiral qBIC samples at a design wavelength of 900 nm approaching the visible wavelength range (Fig. 3). The AFM images (Fig. 3a) highlight the excellent experimental reproduction of the designs with clearly defined structure sizes and height differences (see also Fig. S6). The optical response of the metasurfaces is retrieved using a home-built transmission microscopy setup (see Fig. S7) incorporating the necessary polarizers and quarter wave plates to generate circularly polarized light. Notably, two beam paths for RCP and LCP light are implemented, enabling the convenient switching between left-handed and right-handed illumination during experiments (for details see Methods). Light is condensed on the sample using a 10x objective and collected through an analyzer using a 60x objective to enable the targeted polarization resolved interrogation of individual metasurface patterns with sizes of 40 by 40 µm². A spectrometer is used for spectral analysis in order to resolve the sharp features of the chiral qBIC resonances.

Focusing first on the achiral structures as depicted in Fig. 3b, we observe nearly identical transmission spectra for RCP and LCP illumination, leading to a vanishing $\Delta T$ signal (Fig. 3b, for AFM images see Fig. S8) throughout a large spectral range from 500 nm to 1000 nm. The spectral features at 900 nm in both the RCP and LCP spectra are attributed to a height-driven BIC.

As apparent from Fig. 3c, the chiral left-handed geometry shows a markedly different optical response, with a pronounced transmission dip at the design wavelength of 900 nm for LCP, which is almost completely absent for RCP illumination in line with our numerical predictions (Fig. S9). This distinct chiral selective behavior results in a large modulation of the $\Delta T$ signal with $\Delta T = 0.7$ and a narrow bandwidth with a quality factor of $Q \approx 80$, which is, to the best of our knowledge, the highest experimental value for a strongly chiral optical resonance reported in the literature so far. For the chiral right-handed structure in Fig. 3d, the situation is reversed, with a strong resonance feature for RCP light and a vanishing optical response for LCP, resulting in a $\Delta T$ value with equally large magnitude but opposite sign at $\Delta T = -0.7$, and a similarly high $Q$ factor as obvious from Fig. 3e and Fig. 3f (for full spectra see Fig. S10 and Fig. S11). Most importantly, the cross-polarization measurements in Fig. 3d show no significant polarization conversion as expected from eq. (2) and our numerical predictions (Fig. 1f).

Additionally, the chiral qBIC retains the versatile resonance tuning capabilities of the BIC concept, which we demonstrate by experimentally varying the orientation angle $\theta$ between the two rods. We find that the modulation of the $\Delta T$ signal (Fig. 3f, left-hand side) is mostly constant while the Q factor of the resonance (Fig. 3f, right-hand side) can be tailored via the diverging angle for $\theta \geq 5°$. Specifically, lower values of $\theta$ produce higher $Q$ factors, which can be beneficial for spectrally selective chiral applications, whereas the signal modulation vanishes approaching the achiral case ($\theta = 0°$). Here, the BIC-based resonance tuning can be harnessed to precisely configure the optical system for the target chiral use case, striking a balance between $Q$ factor and modulation as required.

Furthermore, to estimate the accuracy of optical experiments, we check the consequences of Lorentz reciprocity by performing the identical measurements on the chiral metasurface flipped upside down. Since the spectral responses remain unchanged (see Fig. S12) we conclude on the sufficient precision of the optical setup, also confirmed by our in-depth analysis (see Fig. S13).

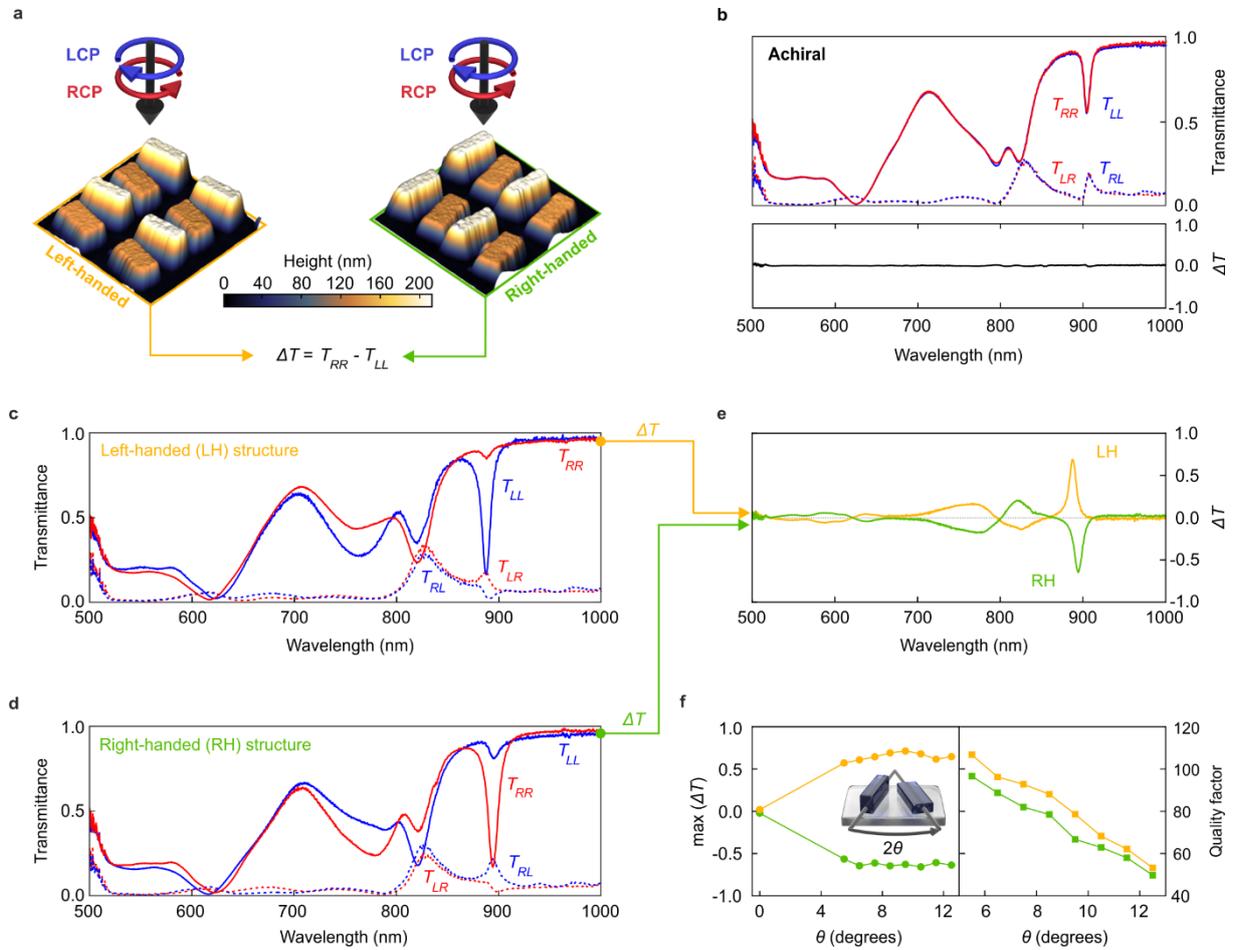

**Fig. 3 | 3D all-dielectric metasurfaces for chiral and spectrally sharp responses. a**, 3D AFM images for left-handed (yellow) and right-handed (green) structures. Both metasurfaces are illuminated with RCP and LCP light and their polarized transmittance coefficients are recorded from which we deduce the chiral transmittance difference $\Delta T = T_{RR} - T_{LL}$. **b**, LCP and RCP transmittance spectra for the achiral metasurface showing almost identical responses in the upper panel. Lower panel: Corresponding $\Delta T$ spectrum showing no chiral selective response. Spectra of LCP/RCP co- and cross-polarized transmittance of the left-handed structure in (**c**) and right-handed structure in (**d**) indicating strongly selective interaction with LCP (**c**) or RCP (**d**) light. **e**, $\Delta T$ spectra for both handednesses show the inversion of the chiral quasi-BIC. **f**, $\Delta T$ peak modulation for different rotation angles $\theta$ between the adjacent resonators for left-handed (yellow) and right-handed (green) structures showing a saturation in the left panel while the extracted $Q$ factors decrease towards higher rotation angles as indicated in the right panel.

**Discussion**

Our experimental polarized transmittance spectra undoubtedly identify the observed resonance at a wavelength of around 900 nm as a chiral qBIC closely tuned to the maximum chiral regime, with the transmittance difference reaching $\Delta T = \pm 0.7$. Thus, for the right-handed (left-handed) enantiomer, the co-polarized LCP (RCP) transmittance stays close to 0.9, while the co-polarized RCP (LCP) transmittance resonantly drops down to below 0.2, while both cross-polarized transmittances remain below 0.1 within the whole range of this resonance.

Qualitatively, regarding the selective RCP/LCP transmission, the qBIC metasurface represents a maximum chiral version of previous intrinsically chiral and rotationally symmetric metasurfaces[39,40] or, in even simpler terms, a maximum chiral analog of a bi-isotropic Pasteur medium[41]. However, while the previous examples absorb the attenuated circularly polarized waves, here, on the contrary, the qBIC metasurface reflects them with up to 80 % efficiency (see Fig. S1). This peculiar type of near-lossless maximum chirality has been previously observed in model microwave experiments[31], and we present its first optical realization.

Until now, selective reflection of circularly polarized light was attributed to the so-called chiral mirrors, which, however, inevitably invert the circular polarization of the transmitted light[42]. Recently, such mirrors have been experimentally realized as perforated silicon slabs[43] and as arrays of silicon particles of reduced symmetry hosting specific qBICs[26]. Being achiral, as all planar metasurfaces, such structures exhibit intriguing optical performance, which, however, elucidates the limits of the planar design and fabrication. Unlocking the out-of-plane dimension enables creating optical structures hosting strongly chiral qBICs selectively coupled to the light of specific helicity. Analogous resonators on a much larger scale can be arranged as Fabry-Perot cavities confined between planar chiral mirrors[44].

In summary, we have suggested and realized a novel nanofabrication approach that allows for the implementation of multi-height level metasurfaces to achieve the precise tailoring of the Q factor in symmetry-broken qBIC metasurfaces via the height difference of the meta-unit resonators. We have revealed how to employ this degree of freedom for the simultaneous tailoring of the qBIC chirality and Q factor, thus opening vast prospects for multi-functional and lossless metasurfaces impacting nonlinear chiral optics and lasing, chirality sensing and, prospectively, enantiomer-selective photochemistry.

**Methods**

**Numerical simulations.** The refractive indexes of the SiO$_2$ substrate and Si rods were taken from the data of in-house white-light ellipsometry. Numerical simulations of the linear height-driven BIC structures were carried out with CST Microwave Studio, determining the spectral responses of height-driven metasurfaces in the frequency domain. Light transmission and reflection by the chiral metasurfaces were numerically investigated by finite-element method (FEM) using the Electromagnetic Waves Frequency Domain module of COMSOL Multiphysics in 3D mode. Tetrahedral spatial mesh for FEM was automatically set by the COMSOL physics-controlled preset. The simulations were performed within a rectangular spatial domain containing one metasurface meta unit with periodic boundary conditions set on its sides and with the excitation and registering circularly polarized ports at the top and the bottom. For Fig. S9, for the wavelength range below the diffraction cutoff, appropriate registering diffraction ports were added. To assess the simulation precision, the transmission and reflection coefficients bound by Lorentz reciprocity were independently obtained and compared, which ensured that the evaluated values are correct with a 1% accuracy. For reproducing experimental spectra, the dimensions of the structures were adjusted to the acquired AFM and SEM data.

**Nanostructure fabrication.** Prior to the fabrication, the silicon dioxide (SiO$_2$) substrates were cleaned via sonication in acetone at 55 °C, rinsed in iso-propanol (IPA), and dried under nitrogen (N$_2$) flux, followed by oxygen (O$_2$) plasma etching to remove organic residues. After the cleaning process, 100 nm of amorphous silicon (a-Si) were deposited onto the SiO$_2$ substrates via plasma-enhanced chemical vapor deposition (PECVD) using a silane (SiH$_4$) precursor gas at a substrate temperature of 250 °C via a PlasmaPro 100 PECVD from Oxford Instruments. The height-driven all-dielectric metasurface fabrication was based on a three-step electron beam lithography (EBL) process. In particular, all EBL processes were performed using an eLINE Plus from Raith Nanofabrication with an acceleration voltage of 30 kV and a 15 μm aperture by patterning a double layer of polymethylmethacrylate (PMMA), a positive-tone electron beam resist, obtained via spin-coating and a subsequent soft-bake step (80 nm of PMMA 950k A2 on top of 100 nm of PMMA 495k A4 from Kayaku Advanced Materials). Subsequent development in a 3:1 IPA:MIBK (methyl isobutyl ketone) solution followed by the electron beam hard mask evaporation and an overnight liftoff using a PMMA remover (Microposit remover 1165) transferred the exposed pattern into the desired nanostructures.

In the first patterning process, a 100 nm thick gold marker system was defined on top of the 100 nm thin a-Si film, which was used for aligning the following two fabrication steps of the metasurfaces with different heights. This marker system enabled a spatial control of the relative nanostructure position between the two fabrication steps of ≤ 5 nm. In the second patterning run, only the lower elements of the meta units were written and developed followed by the deposition of a 25 nm chromium (Cr) on 20 nm $SiO_2$ hard mask. Subsequent deposition of an additional a-Si layer (here 10 nm, 20 nm, and 40 nm) homogeneously covered the sample and buried all structures beneath. The third lithography step was used to define the taller elements within the meta unit, again with a 25 nm Cr mask on top of 25 nm $SiO_2$.

Finally, the hard mask pattern was transferred into the silicon via reactive ion etching (RIE) using a PlasmaPro 100 ICP-RIE from Oxford Instruments in a mixture of 7 sccm Ar and 20 sccm $Cl_2$ at 20 W bias power and 200 W induction power. After etching, the chromium hard mask was removed in a wet etchant to retrieve the silicon structures. The $SiO_2$ capping layer was removed in another RIE step using 30 sccm Ar and 20 sccm $CHF_3$ under 30 W bias power and the sample was cleaned in gentle oxygen plasma before the optical measurements.

**Optical characterization of height-driven qBICs.** The optical characterization of the achiral height-driven qBIC metasurfaces was performed in a commercially available white light confocal microscopy setup (Witec alpha 300 series). The sample was illuminated with linearly polarized and collimated white light from the bottom by a broadband halogen lamp (Thorlabs OSL2). Afterwards, the transmitted signal was collected with a 10x (NA = 0.25) objective and confocally detected by coupling it into a multimode fiber. From there, the light was guided into the spectrometer where it impinged onto a reflective grating (150 grooves/mm) and was dispersed onto a silicon CCD. For each spectrum, the transmitted intensity through the qBIC metasurface was normalized to the bare $SiO_2$ substrate to retrieve the transmittance spectrum and remove any unwanted features from the optical beam path. For the acquisition of an individual spectrum, 40 accumulations were taken at 0.5 seconds each.

**Optical characterization of the chiral qBICs.** Optical characterization was performed using a custom-built transmission microscope on an optical table. A schematic overview over the set-up is given in Figure S7a. The setup is driven by the collimated output of a fiber-coupled supercontinuum white light laser (SuperK FIANIUM from NKT Photonics) set to a power of 5 % of the maximum value and a repetition rate of 0.302 MHz. The beam was guided to a polarizing beam splitter (PBS), where the beam is split in two paths, generating horizontal (HP) or vertical (VP) linear polarization (2x LPVIS100 from ThorLabs, 550-1500 nm), respectively. Together with the quarter wave plate (QWP, RAC4.4.20 from B-Halle, 500-900 nm), circularly polarized light (CPL) was generated. The polarization could be varied conveniently between RCP and LCP by blocking one of the two beam paths generating HP or VP. This eliminates the need to rotate the polarizers or QWP, which can induce elliptical polarization if not controlled after rotation. The QWP was located directly below the objectives to avoid any reflections off mirrors, which can turn CPL into elliptically polarized light while linearly polarized light remains unaffected.

Light was condensed on the sample using a 10x objective (Olympus PLN, NA = 0.25) and collected using a 60x objective (Nikon MRH08630, NA = 0.7). After alignment, the beam was slightly defocused to illuminate the entire metasurface area of 40 µm x 40 µm (Figure S7b). The desired location for the measurement was selected by using an aperture after the objectives. For the measurement of the co- and cross-polarization terms a chiral analyzer, consisting of a QWP (AQWP05-580 from Thorlabs, 350-850 nm) and a linear polarizer (WP25M-UB from Thorlabs, 250-4000 nm), was installed after the collection objective. Using a flip mirror, the light was then guided directly to the CCD camera or to the spectrometer by using a multimode fiber (Thorlabs M15L05, core size: 105 µm, NA = 0.22). A spectrometer from Princeton Instruments was used with a grating period of 300 g/mm, blaze angle of 750 nm and spectral resolution of 0.13 nm. All spectra were recorded using a binning of one line, an exposure duration of 50 ms and 50 spectra were accumulated. Prior to measurement, the dark noise was recorded by blocking the illumination. All spectra were referenced by using a background measurement, which was performed on the same substrate far away from the metasurfaces.

**Data availability**

The data that support the findings of this study are available from the corresponding author upon reasonable request.

**Acknowledgements**

This work was funded by the Deutsche Forschungsgemeinschaft (DFG, German Research Foundation) under grant numbers EXC 2089/1 – 390776260 (Germany's Excellence Strategy) and TI 1063/1 (Emmy Noether Program), the Bavarian program Solar Energies Go Hybrid (SolTech), and the Center for NanoScience (CeNS). S.A. Maier additionally acknowledges the EPSRC (EP/W017075/1), the Australian Research Council, and the Lee-Lucas Chair in Physics. The work of M.V.G. and A.A.A. was supported by the Ministry of Science and Higher Education of the Russian Federation within the State assignment of FSRC "Crystallography and Photonics" RAS. Y.K. acknowledges a support from the Australian Research Council (grant DP210101292), as well as the International Technology Center Indo-Pacific (ITC IPAC) and Army Research Office under Contract No. FA520921P0034

**Author contributions**

L.K., A.T., Y.K., and M.V.G. conceived the project idea. M.V.G. and A.A.A. developed the theoretical background. L.K. and A.A.A. performed numerical simulations. L.K. fabricated the samples. L.K., F.W., J.B., L.H., and L.S.M. performed the optical measurements. All authors contributed to the data analysis and to the writing of the manuscript.

**Competing interests**

The authors declare no competing interests.

**Additional information**

Correspondence and requests for materials should be addressed to A.T.

# Supporting Information for

# Unlocking the out-of-plane dimension for photonic bound states in the continuum to achieve maximum optical chirality

Lucca Kühner[1,†], Fedja J. Wendisch[1,†], Alexander A. Antonov[2], Johannes Bürger[1],

Ludwig Hüttenhofer[1], Leonardo de S. Menezes[1,3], Stefan A. Maier[4,1,5], Maxim V. Gorkunov[2],

Yuri S. Kivshar[6], and Andreas Tittl[1,*]

[1]Chair in Hybrid Nanosystems, Nanoinstitute Munich, and Center for NanoScience, Faculty of Physics, Ludwig-Maximilians-University Munich, Königinstrasse 10, 80539 München, Germany

[2]Shubnikov Institute of Crystallography, FSRC "Crystallography and Photonics", Russian Academy of Sciences, Moscow 119333, Russia

[3]Departamento de Física, Universidade Federal de Pernambuco, 50670-901 Recife, Pernambuco, Brazil

[4]School of Physics and Astronomy, Monash University, Wellington Rd, Clayton VIC 3800, Australia

[5]The Blackett Laboratory, Department of Physics, Imperial College London, London, SW7 2AZ, United Kingdom

[6]Nonlinear Physics Centre, Research School of Physics Australian National University, Canberra ACT 2601, Australia

†These authors contributed equally

*e-mail: andreas.tittl@physik.uni-muenchen.de

# Contents

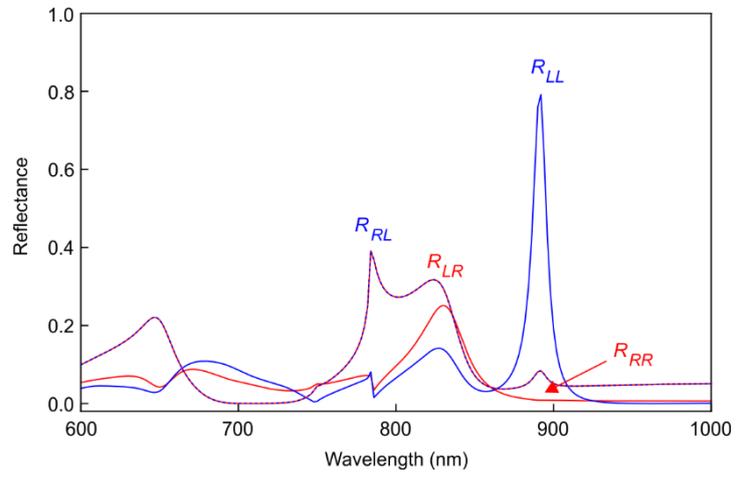

**Figure S1: Simulated reflectance spectra of a left-handed chiral qBIC enantiomer**.

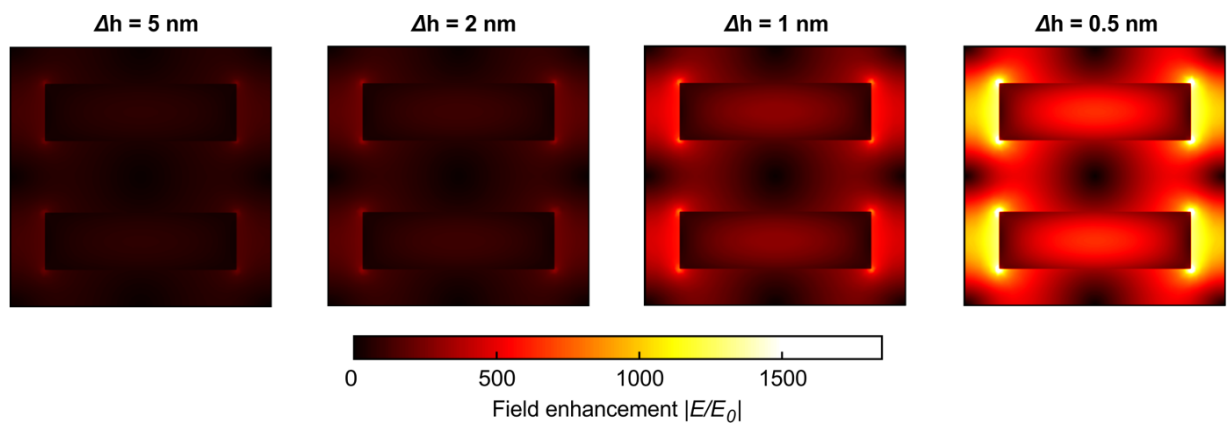

**Figure S2: Angstrom-level control for high field enhancements.** Numerical investigation of the electric near-fields associated with the height-driven qBIC resonances for various small height differences. The field enhancement increases drastically towards Angstrom-scale height differences, such as 5 Å on the right-hand side with field enhancements of up to 2000.

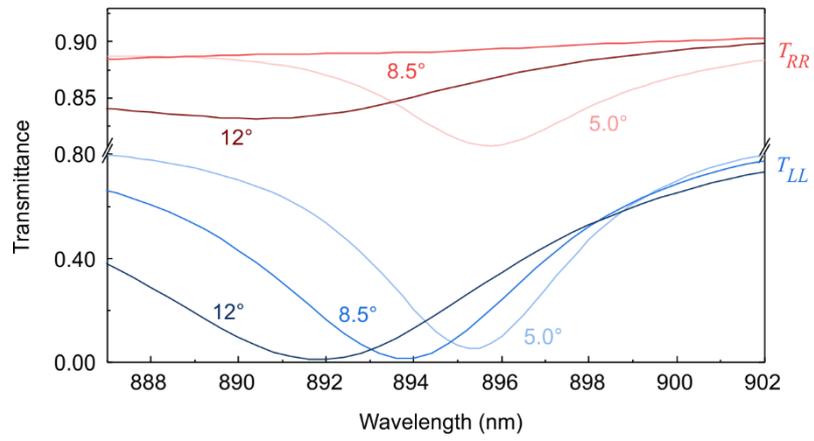

**Figure S3: Simulated transmittance spectra for different diverging angles $\theta$.**

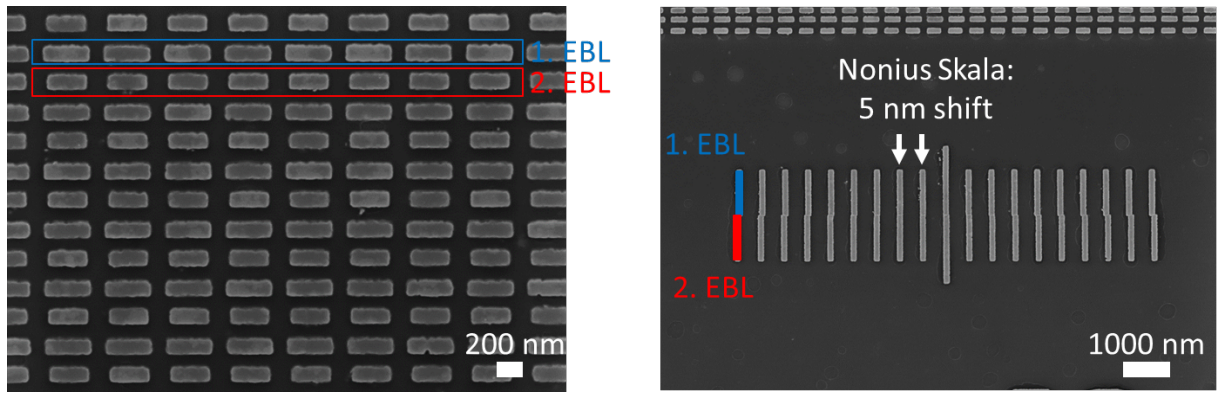

**Figure S4: Lateral resolution of the multi-step nanofabrication approach.** SEM images proving the lateral resolution of the multi-step nanofabrication approach. Left: Fabricated metasurfaces with two electron beam lithography (EBL) steps, where both resonators have the same height. Right: Nonius scale for determining the shift between both EBL steps. The lines at the top and bottom have been written with two EBL steps (marked blue and red). From the center (largest line), each line is shifted by $\pm$ 5 nm for the second EBL step, revealing a spatial resolution of < 5 nm, which can be taken into account during the sample fabrication.

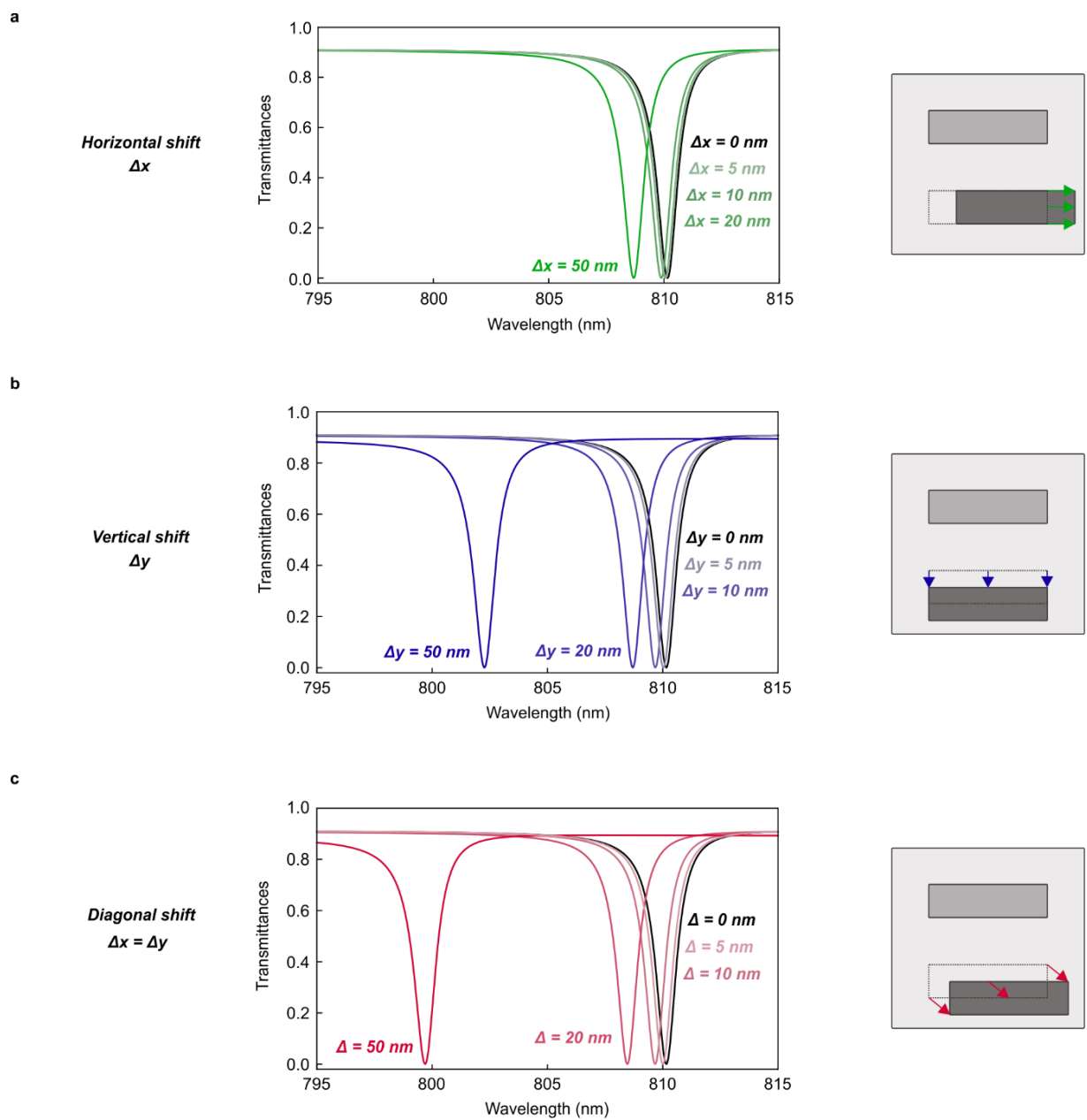

**Figure S5: Numerical investigation of the influence of misalignment on the height-driven BIC spectra.** Simulated spectral responses for small shifts in *x* (panel a), in *y* (panel b), and in diagonal direction (panel c) show no significant resonance shifts for alignment accuracies below 10 nm. The properties of the resonance also remain mostly unchanged.

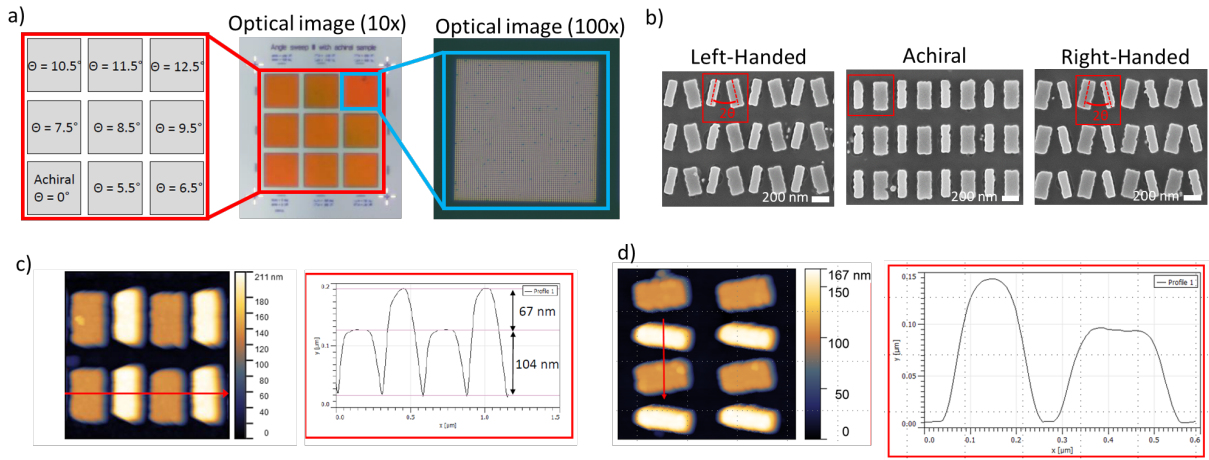

**Figure S6: Fabrication and characterization of the achiral and chiral qBIC metasurfaces. a,** Left: Overview over the different achiral and chiral qBIC metasurfaces with varying opening angles. Middle: Optical image (10x) of the fabricated qBIC metasurfaces. Right: Optical image (100x) of the chiral qBIC metasurfaces with opening angle of 12.5°. **b,** SEM images of the achiral and chiral left- and right-handed sample. The meta-atom with larger height is visible by its brighter contrast. **c, d,** AFM measurements and line scans of the achiral (**c**) and chiral (**d**) qBIC metasurface.

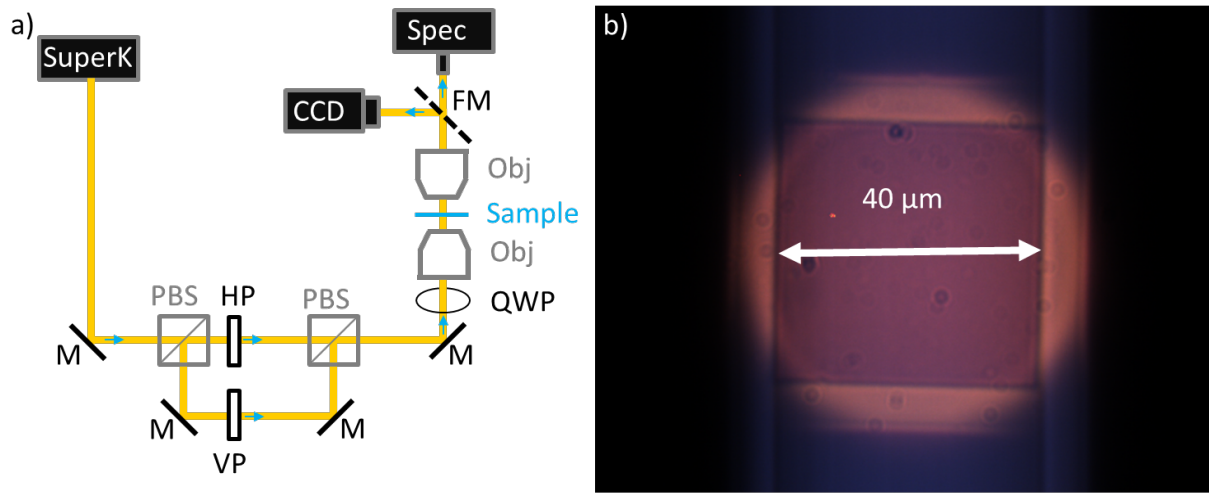

**Figure S7: Overview over the set-up for optical characterization of the chiral qBICs. a,** Schematic overview of the optical characterization setup. M (solid black line): mirror, FM (dashed black line): flip mirror. PBS: polarizing beam splitter. HP: horizontal linear polarizer. VP: vertical linear polarizer. QWP: quarter wave plate. Obj.: microscope objective. To generate RCP or LCP light, either HP or VP beam path is blocked. **b,** Image of the qBIC metasurface captured on the SSD camera.

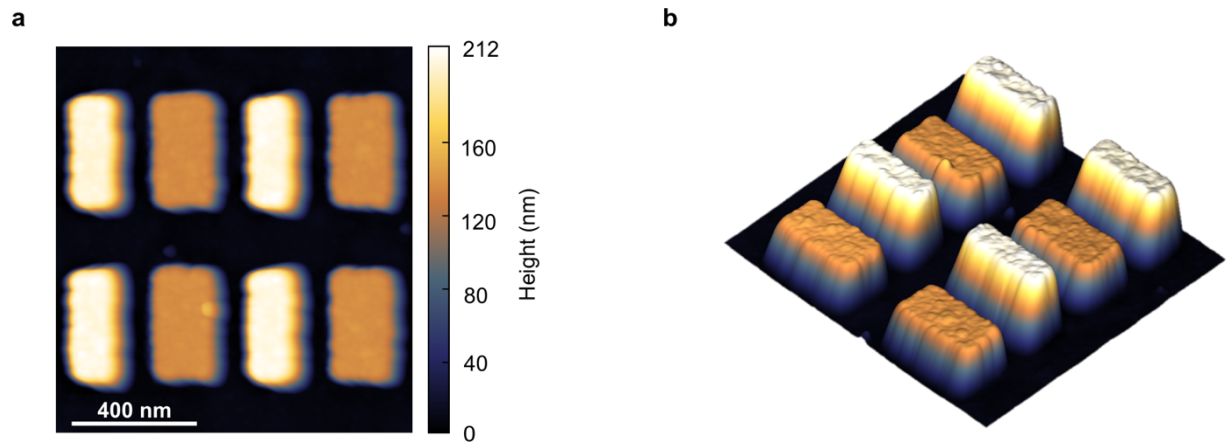

**Figure S8: AFM images of the achiral qBIC metasurface.** Planar (**a**) and 3-D (**b**) AFM image for the achiral qBIC metasurface.

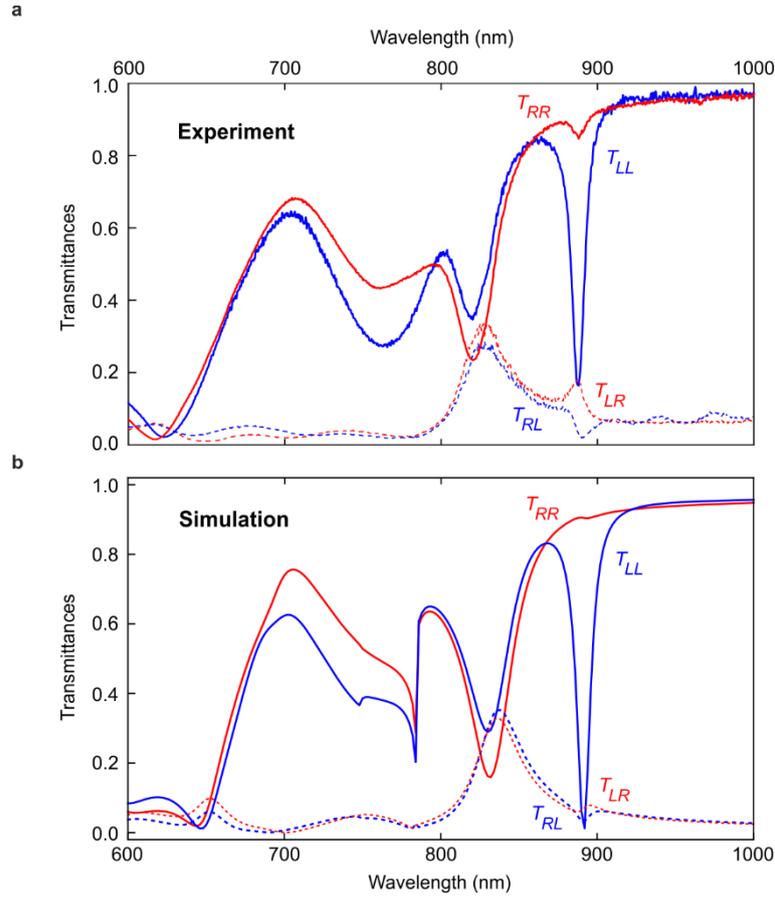

**Figure S9: Comparison between experiment and simulation.** The measured chiral response (upper panel) for a left-handed structure with an opening angle of $\theta = 8.5°$ only interacts with LCP light (blue curve $T_{LL}$) for which the chiral qBIC is excited. The experimental data show excellent agreement with our numerical simulations (lower panel). We used scanning electron microscopy (SEM) images to match the lateral dimensions, atomic force microscopy scans (see Fig. S6 for SEM and AFM data) to determine the different heights of the resonators and utilized the in-house measured amorphous silicon ellipsometry data to match the numerical simulations as precisely as possible to the measurement results. In particular, the simulations were done for a periodicity of 550 nm, heights of $h_1 = 104$ nm and $h_2 = 171$ nm (see Fig. S6), an opening angle of $\theta = 8.5°$, rod widths of $w_1 = 104$ nm and $w_2 = 157$ nm, and lengths of $l_1 = l_2 = 334$ nm.

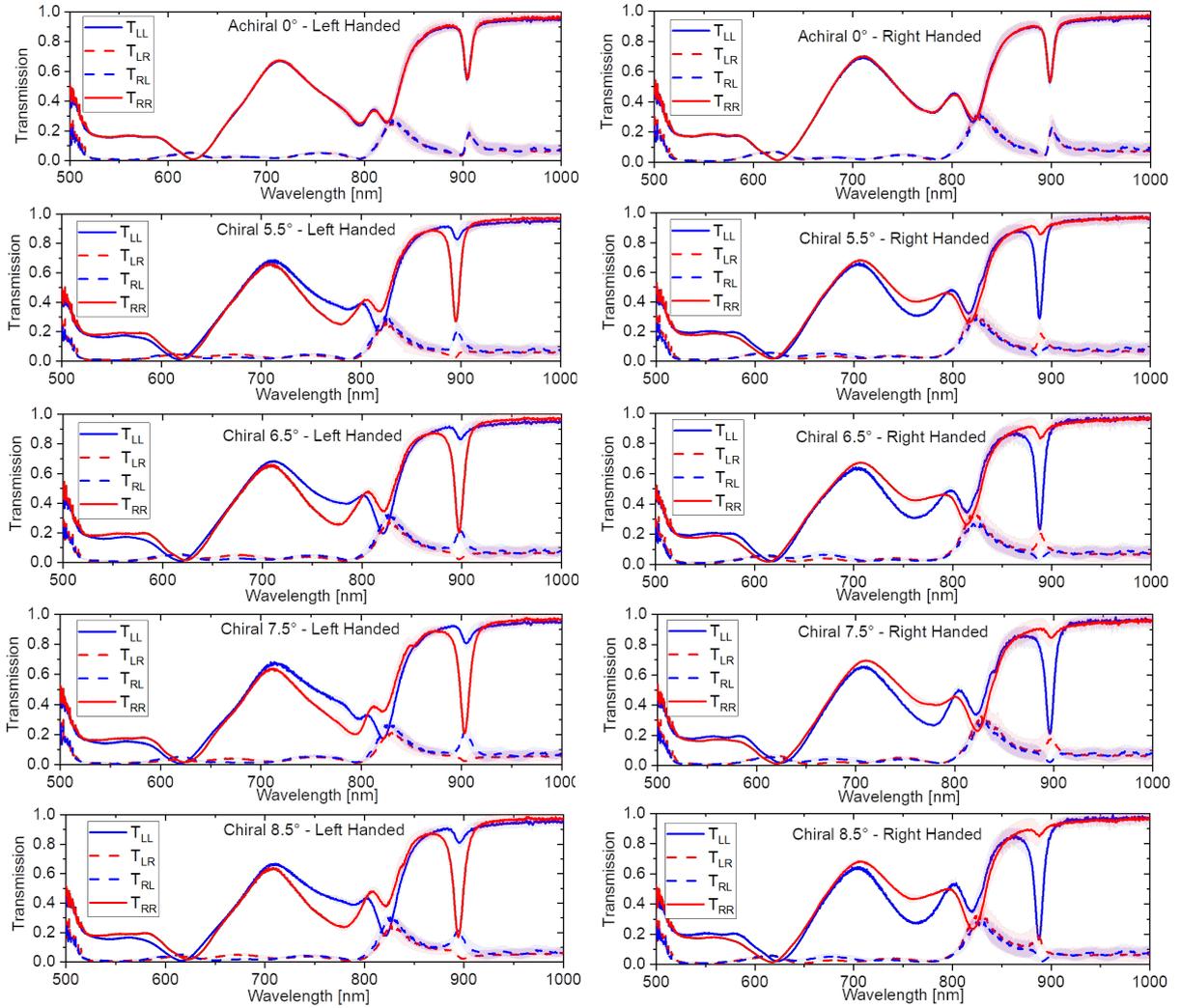

**Figure S10: All experimental spectra of the achiral and chiral qBIC metasurfaces with opening angles from 0-8.5°.** Left: Left-handed structures. Right: Right-handed structures. The opening angle increases from top to bottom. All metasurfaces were measured with a chiral analyzer. The solid blue and red lines represent the co-polarization terms, $T_{LL}$ and $T_{RR}$, respectively, whereas the dashed blue and red lines represent the cross-polarization terms, $T_{LR}$ and $T_{RL}$, respectively. The solid and dashed lines are the median of 4 measurements (see Figure S13 for more details). The shaded area represents the standard deviation of these 4 measurements.

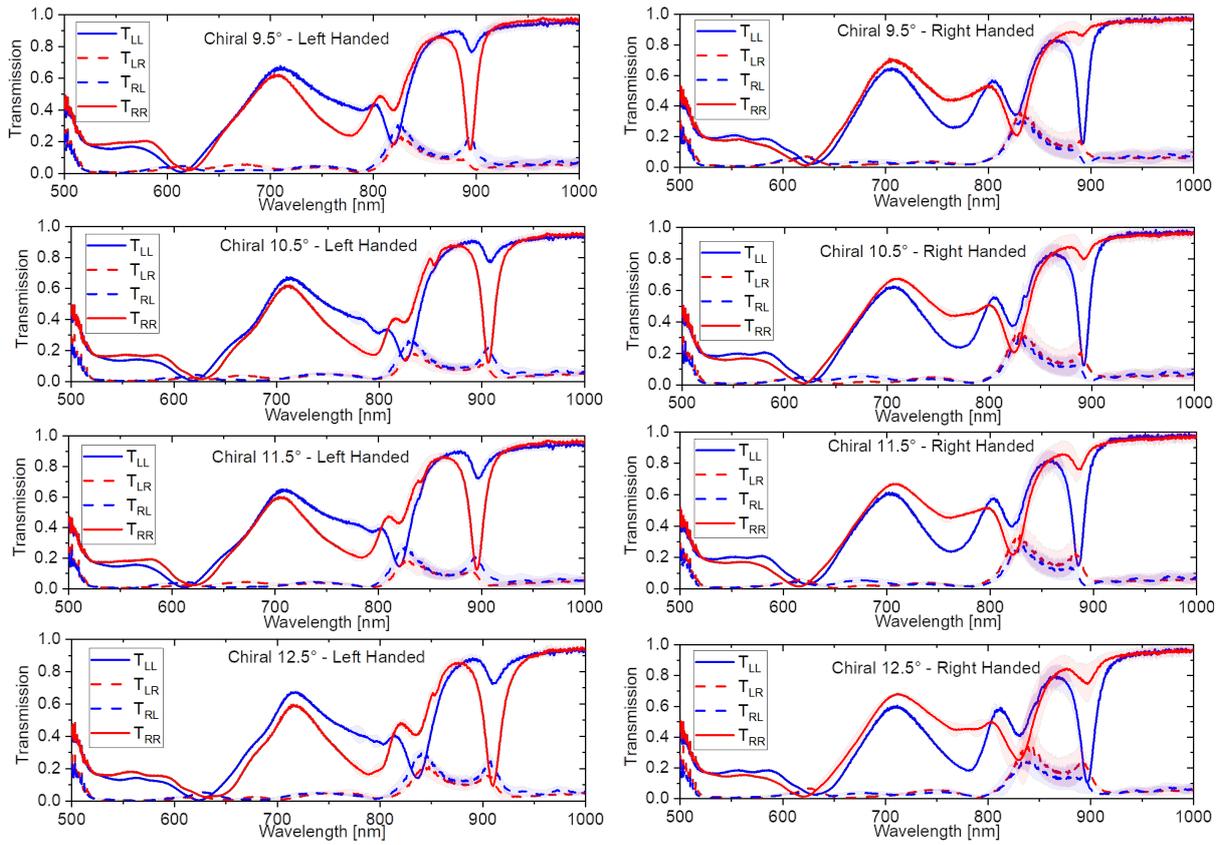

**Figure S11: All experimental spectra of the achiral and chiral qBIC metasurfaces with opening angles from 9.5-12.5°.** Left: Left-handed structures. Right: Right-handed structures. The opening angle increases from top to bottom. All metasurfaces were measured with a chiral analyzer. The solid blue and red lines represent the co-polarization terms, $T_{LL}$ and $T_{RR}$, respectively, whereas the dashed blue and red lines represent the cross-polarization terms, $T_{LR}$ and $T_{RL}$, respectively. The solid and dashed lines are the median of 4 measurements (see Figure S13 for more details). The shaded area represents the standard deviation of these 4 measurements.

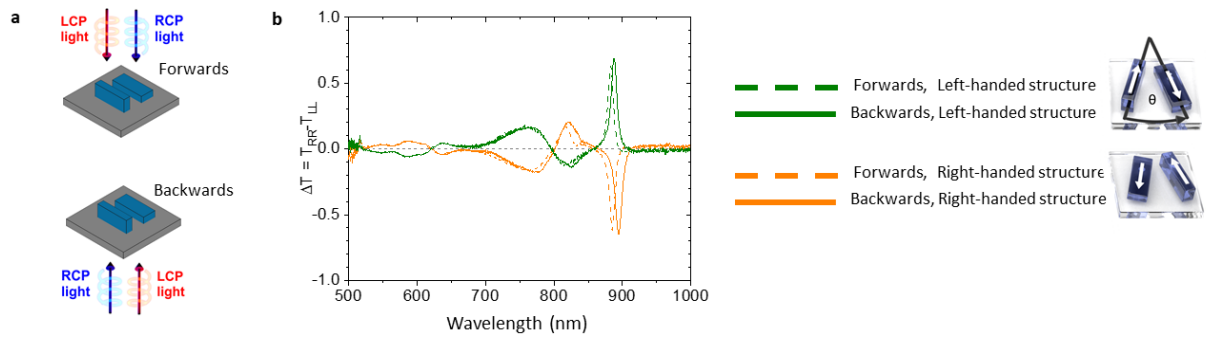

**Figure S12: Lorentz reciprocity of the chiral qBIC. a,** Schematic illustration showing the flipping of the substrate to measure it in forwards (solid line) and backwards (dashed line) direction. **b,** Transmittance difference for the right-handed (orange) and left-handed (green) qBIC metasurfaces. The transmittance difference signal of the qBIC metasurfaces does not flip due to their three-dimensionality, contrary to planar chiral structures.

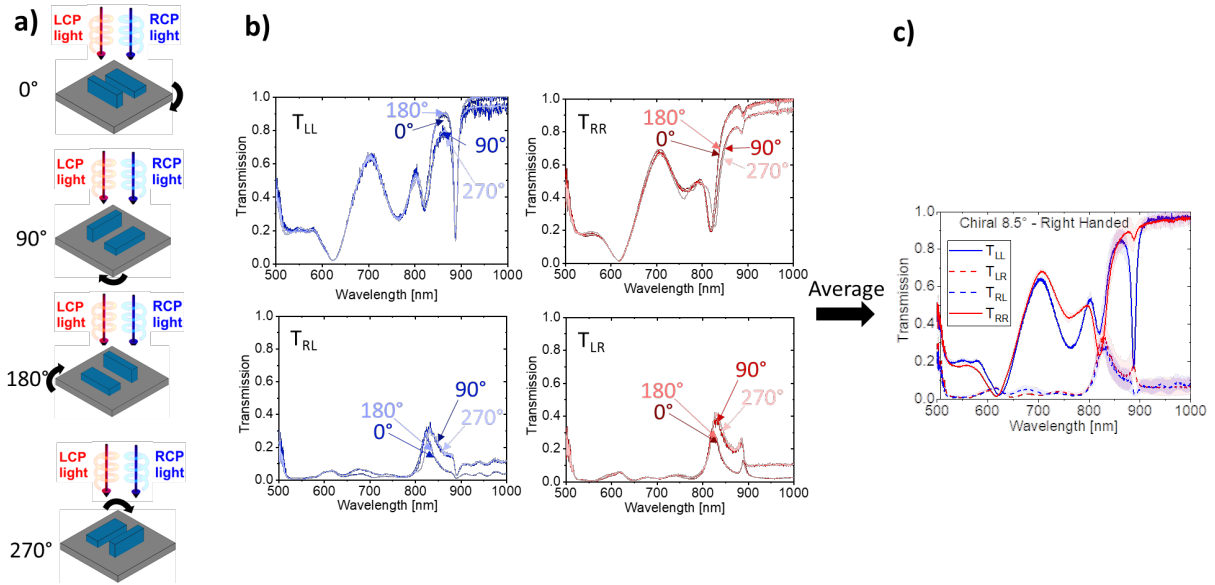

**Figure S13: Influence of sample rotation on the measurements of the chiral qBIC metasurfaces with chiral analyzer. a,** Schematic overview of the performed sample rotations during measurement. The sample was rotated four times with a 90° angle to avoid any influences of elliptical polarization or sample tilt. **b, c,** Single spectra of the four measurements (**b**) and the mean value with standard deviation (**c**) of the right-handed chiral qBIC metasurfaces with an opening angle of $\theta = 8.5°$ measured with chiral analyzer. A small signal variation in the transmittance data upon sample rotation is evident in the four single spectra for each co- and cross-polarization ($T_{LL}$, $T_{RR}$, $T_{LR}$, $T_{RL}$). The signal is identical for rotations of 180° (0°, 180° and 90°, 270°), which is a hint towards a small degree of elliptical polarization, which decreases the measured CD. For an influence of sample tilt, the signal should vary for all rotation angles. The small degree of elliptical polarization might be caused by the deviation of retardation (+/- 6%) given by the quarter wave plate (QWP, RAC4.4.20 from B-Halle, 500-900 nm) and by deviation of the chiral analyzer (QWP, AQWP05-580 from Thorlabs, 350-850 nm; polarizer WP25M-UB from Thorlabs, 250-4000 nm).